\documentclass[10pt,letterpaper,twocolumn]{article} %% two column, final layout

\usepackage{ol2}
\usepackage[draft]{hyperref}
\usepackage{amsmath}
\usepackage[english]{babel}
\begin{document}

\twocolumn[ %% activate for two-column option

\title{Experimental demonstration of a dual-frequency laser free from anti-phase noise}

\author{Abdelkrim El Amili,$^{1,*}$ Goulc'hen Loas,$^1$ Syamsundar De,$^2$ Sylvain Schwartz,$^3$ Gilles Feugnet,$^3$ Jean-Paul Pocholle,$^3$ Fabien Bretenaker,$^2$ and Mehdi Alouini$^{1,3}$}

\address{
$^1$Institut de Physique de Rennes, Universit\'e de Rennes 1, CNRS, Campus de Beaulieu, 35042 Rennes, France
\\
$^2$Laboratoire Aim\'e-Cotton, CNRS-Universit\'e Paris 11, 91405 Orsay Cedex, France\\
$^3$Thales Research and Technology, RD 128, 91767 Palaiseau Cedex, France\\
$^*$Corresponding author: abdelkrim.elamili@univ-rennes1.fr
}

\begin{abstract}
A reduction of more than 20 dB of the intensity noise at the anti-phase relaxation oscillation frequency is experimentally demonstrated in a two-polarization dual-frequency solid-state laser without any optical or electronic feedback loop.  Such a behavior is inherently obtained by aligning the two orthogonally polarized oscillating modes with the crystallographic axes of a (100)-cut neodymium-doped yttrium aluminum garnet active medium. The anti-phase noise level is shown to increase as soon as one departs from this peculiar configuration, evidencing the predominant role of the nonlinear coupling constant. This experimental demonstration opens new perspectives on the design and realization of extremely low noise dual-frequency solid-state lasers.

\end{abstract}

 ] %% activate for two-column option

Dual-frequency solid-state lasers are attractive for a large number of applications such as microwave photonics \cite{Alouini2001, Brunel2005,Pillet2008}, spectroscopy \cite{Brunel1997}, and metrology \cite{Nerin1997, Du2005}. In particular, when the two modes are cross-polarized, dual-frequency lasers are shown to be well suited for obtaining large tunability of the frequency difference, voltage controlled tunability, as well as compactness  \cite{Pillet2008,McKay2009}. In this context, different gain media, either crystals or glasses, may be used to reach different wavelengths \cite{Brunel1997,Alouini1998, Czarny2004,Wang2009}. While such solid-state lasers are known to exhibit very narrow spectral widths, they suffer from resonant intensity noise at low frequencies, i.e., from a few kHz to a few MHz \cite{Alouini2001}. As far as dual-frequency lasers are considered,  the intensity noise spectrum of each eigenmode exhibits two peaks lying at the well know in-phase and anti-phase eigen-frequencies of two coupled oscillators \cite{Otsuka1992}. The in-phase noise, which corresponds to the standard relaxation oscillations of the laser, can be reduced either electronically or optically using feedback loops \cite{Taccheo1996,Alouini2001}. However, the anti-phase noise, which is related to a resonant exchange of energy between the two laser modes, is very difficult to circumvent \cite{Pillet2008} because the reduction of this noise would require an additional servo-loop acting on the difference of the intensities of the two modes or two servo-loops acting independently on the intensity of each  mode. The existence of the anti-phase noise being by essence due to the fact that the two laser modes share totally or partially the same population inversion, another approach is to separate spatially the two lasers modes in the active medium \cite{Alouini1998,Yifei2001,Czarny2004}. Nevertheless such a two axis approach increases the complexity of the laser. Moreover it reduces the correlation between the frequency jitter of the two modes as compared to a single axis, thus reducing the efficiency of the common mode noise rejection. Besides, another solution consists in using class-A lasers \cite{Baili2009}, which are free from relaxation oscillation. But this is usually not possible for solid-state lasers. Consequently, an optimal dual-frequency laser in terms of intensity noise and beat frequency stability would be a single axis laser in which the population inversions related to each mode are independent.

In this Letter, we experimentally demonstrate how the proper design of a two polarization dual-frequency solid-state laser allows to get rid of the anti-phase noise in the simplest possible architecture and without using any electronic or optical feedback loop. This design is based on an appropriate choice of the active medium cut and orientation in order to assign two almost independent families of active atoms to the two laser modes.
\begin{figure}[htb]
\centerline{
\includegraphics[width=7cm]{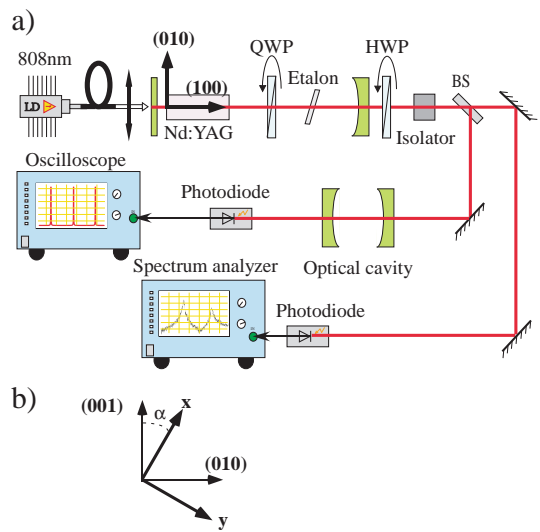}}
 \caption{a) Experimental setup. QWP: quarter-wave plate; HWP: half-wave plate; BS: beam splitter. b) $\alpha$ is the angle between the $x$ and $y$ eigenpolarization directions and the Nd:YAG crystallographic axes.}
 \label{Fig1}
\end{figure}
The simple two-frequency laser architecture that we chose is schematized in Fig. \ref{Fig1}(a). The two-mirror cavity contains a quarter-wave plate (QWP) which defines the orientations of the two eigenpolarizations of the laser. In this case these two eigenpolarizations are linear and aligned along the neutral axes of the QWP. Their frequency difference is equal to one half of the free spectral range of the cavity. Of course, in usual two-frequency lasers \cite{Brunel1997}, one uses a variable intracavity retardance in order to be able to tune the frequency difference between the two modes. Here we restrict to the simple case of an intracavity QWP because rotating this QWP permits to simply rotate the orientation of the eigenpolarizations. The question now is how should we choose our active medium in order to uncouple the two polarization modes, i. e., in order to minimize cross-saturation effects among our two modes ? It has recently been shown that, in Nd:YAG, the emitting dipoles behave as if they were aligned along the crystallographic axes of the matrix \cite{Schwartz2009}. In particular, by choosing a (100)-cut Nd:YAG crystal instead of the more common (111) cut, it was shown that almost complete decoupling of two perpendicularly polarized modes could be obtained by aligning them with the (010) and (001) crystallographic axes. We have thus chosen to use such a crystal here and to observe the evolution of the laser intensity noise spectrum when we rotate the QWP with respect to the crystal axes. As depicted in Fig.\ref{Fig1}(a), the laser is based on a 7-cm-long planar-concave cavity. The gain medium is a 2-cm-long (100)-cut Nd:YAG crystal. The crystallographic axes were precisely determined by X-ray diffraction. In order to limit thermally induced birefringence, the crystal is placed inside a copper mount. It is pumped by a cw multimode fiber-coupled low power laser diode (300 mW) operating at 808 nm. We have checked that the pump beam is thus depolarized, avoiding any pump induced gain anisotropy \cite{Bouwmans2001}. The QWP defining the two eigenpolarization directions $x$ and $y$ is mounted on a precise rotation mount in order to control the angle $\alpha$ between the Nd:YAG crystallographic axes and the polarization states (Fig. \ref{Fig1}(b)). An intra-cavity silica \'etalon forces the laser to oscillate in a single longitudinal mode for each polarization state. Both modes are continuously analyzed with a Fabry-P\'erot cavity to check that the laser remains monomode without any mode hop during data acquisition. Laser output is detected using a photodiode (3.7 MHz bandwidth). A half-wave plate followed by an isolator in front of the detector permit to project the laser output on any linear polarization state before detection by rotating the half-wave plate. The noise spectrum is recorded with an electrical spectrum analyzer (ESA).
\begin{figure}[htb]
\centerline{
\includegraphics[width=8.cm]{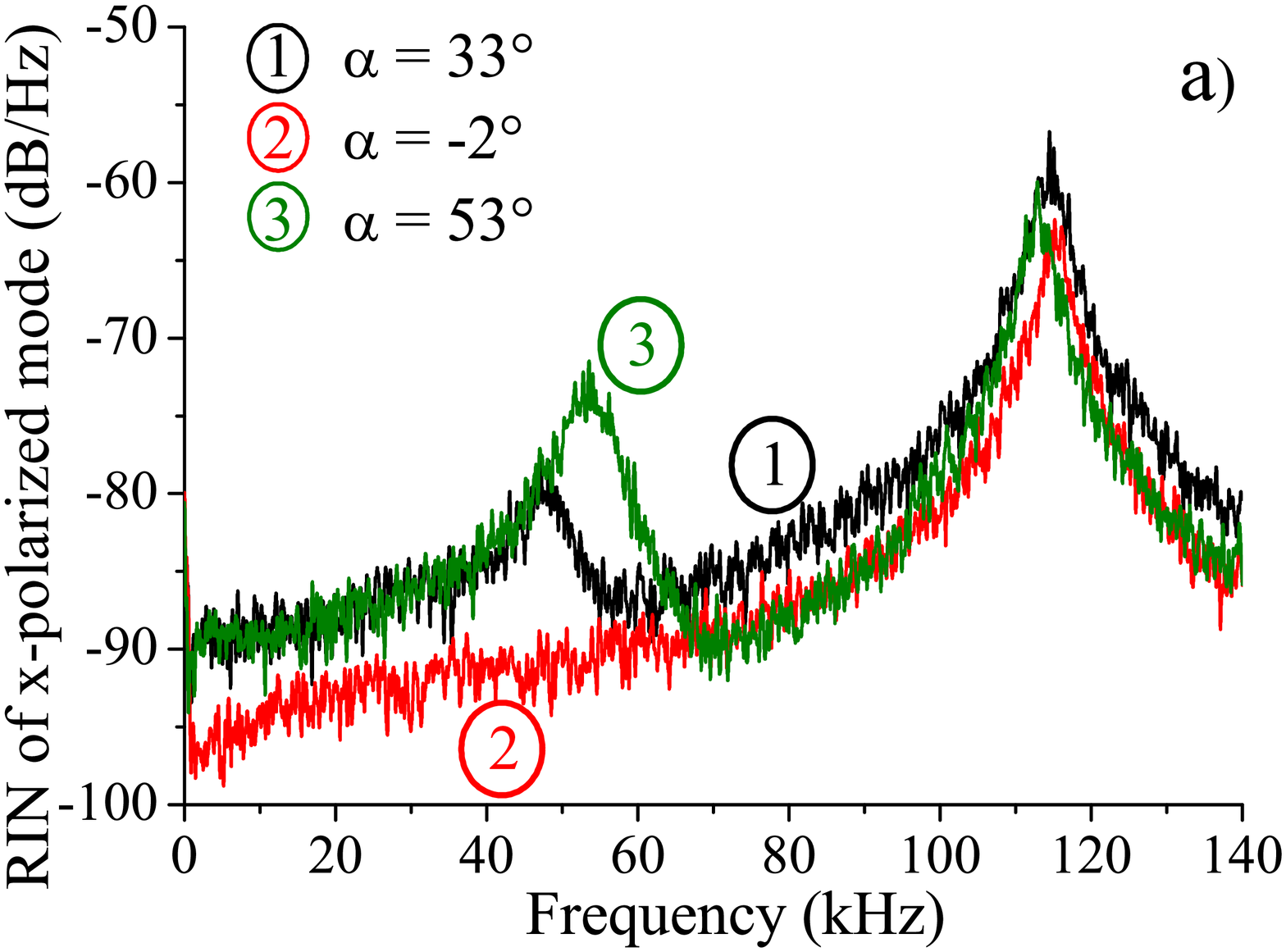}}
\centerline{
\includegraphics[width=8.cm]{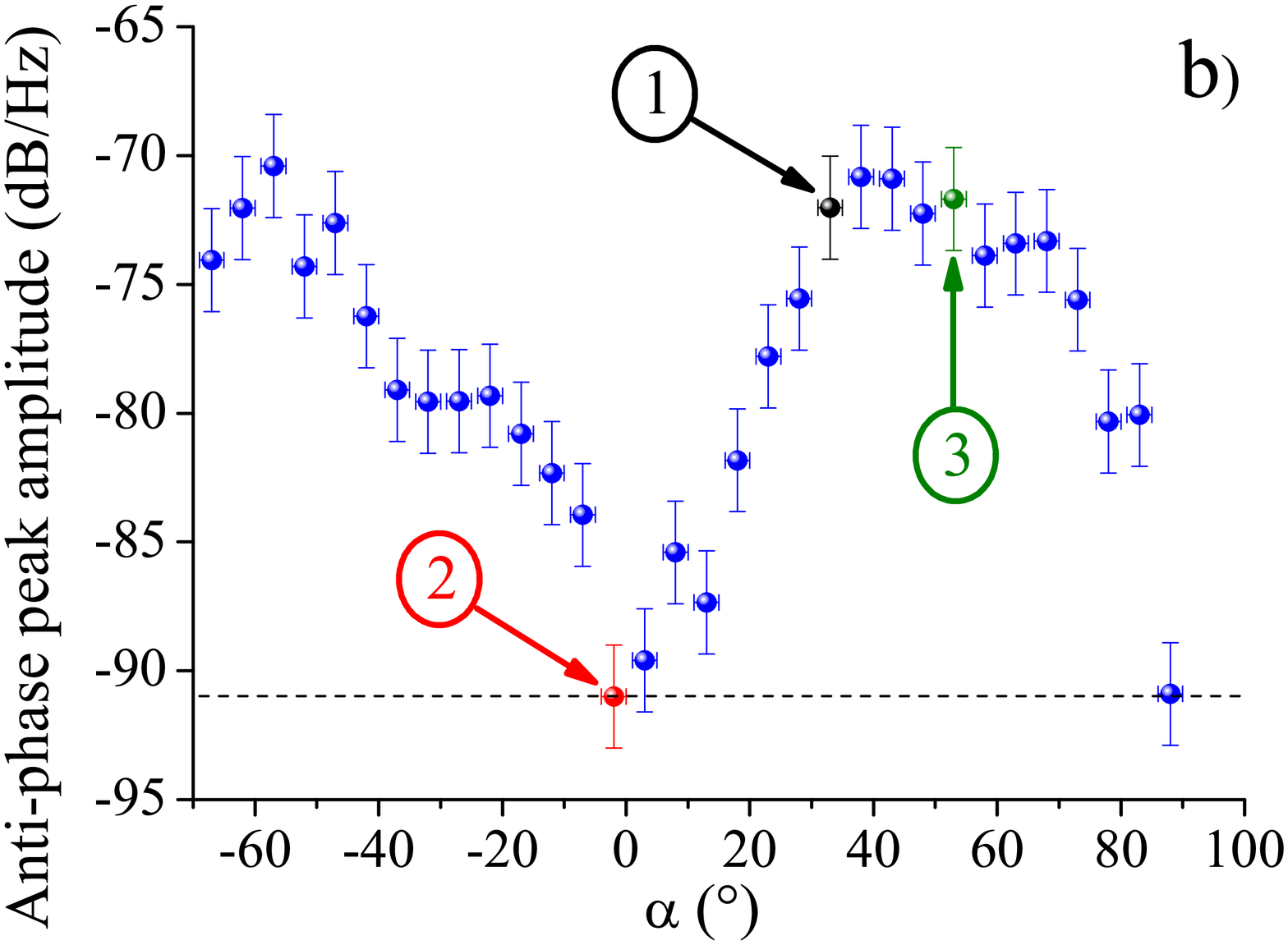}}
 \caption{a) RIN spectra for the $x$-polarized mode only for different values of $\alpha$ (ESA resolution bandwidth: 200 Hz; video bandwidth: 200 Hz). Note that the anti-phase peak has disappeared for $\alpha = -2^{\circ}$. b) Evolution of the amplitude of the anti-phase peak versus $\alpha$. Dashed line: noise floor level below which the anti-phase peak is no longer measurable. The pump power is 300~mW and the laser is 1.5 times above threshold. The laser's output power is 20~mW.}
\label{Fig2}
\end{figure}
Fig. \ref{Fig2}(a) reproduces the relative intensity noise (RIN) spectra recorded when only the $x$-polarized mode is detected for three values of $\alpha$. For $\alpha=33\,^{\circ}$ and $\alpha=53\,^{\circ}$, we observe the existence of the usual in-phase relaxation oscillation peak at 115~kHz and of the anti-phase relaxation oscillation peak at about 50 to 60~kHz. Of course, while the in-phase peak is always there, the anti-phase peak can be hidden when one balances the intensities of the two modes on the detector. Now, we rotate $\alpha$ while detecting only the $x$ polarization and look for a position in which the amplitude of the anti-phase peak is minimized. This leads to $\alpha=-2\,^{\circ}$, and to the red spectrum in Fig. \ref{Fig2}(a). As expected, for this orientation which is close to $\alpha=0$ for which the coupling is expected to be minimum \cite{Schwartz2009}, the anti-phase peak becomes so small that it disappears below the noise floor. The evolution of the anti-phase peak amplitude versus $\alpha$ is plotted in Fig. \ref{Fig2}(b). One can see that this amplitude is maximum (resp. minimum) for $\alpha$ close to $\pm \pi/4$ (resp. $0$ or $\pi/2$), i. e., when the coupling is expected to be maximum (resp. minimum). This is consistent with the fact that the laser behaves as if the emitting dipoles were aligned along the crystallographic axes, like in Ref. \cite{Schwartz2009}.
\begin{figure}[htb]
\centerline{
\includegraphics[width=8.cm]{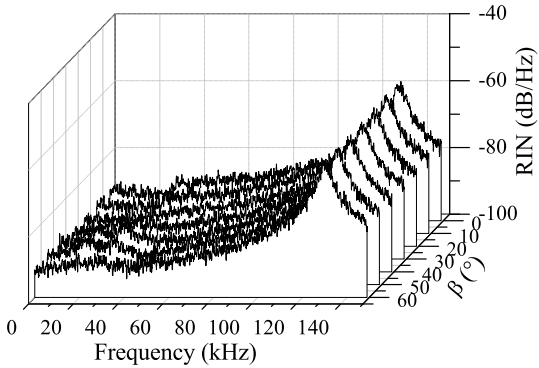}}
 \caption{RIN spectra versus HWP orientation $\beta$ for $\alpha=88\,^{\circ}$. ESA resolution bandwidth: 200 Hz; video bandwidth: 200 Hz. }
 \label{Fig3}
\end{figure}
In the interesting situations corresponding to $\alpha=-2\,^{\circ}$ and $\alpha=88\,^{\circ}$ in which the anti-phase relaxation oscillation noise peak has been minimized, we expect this peak to be disappeared for all orientations of the polarization analyzer located in front of the detector.  This is what we check in Fig. \ref{Fig3}, which shows the RIN spectra obtained for several orientations $\beta$ between $x$ and the fast axis of the HWP. The HWP is rotated by steps of $10^{\circ}$. This corresponds to a rotation of the passing axis of the polarization analyzer by $2\beta$ with respect to $x$. These experimental spectra show, on the one hand, that the RIN behavior remains almost the same for all orientations of the polarization analyzer, and, on the other hand, that the anti-phase peak is drastically reduced for both polarization modes of the laser.

In summary, we have shown that the noise induced by the anti-phase relaxation oscillation resonance in a dual-frequency laser can be almost completely cancelled by a proper choice of the orientation of the laser eigenpolarizations with respect to the orientations of the light emitting dipoles. This has been illustrated in the case of a Nd:YAG crystal in which the choice of a (100)-cut crystal together with the proper orientation of the polarizations of the laser modes permits to cancel this resonance by more than 20 dB. This is in perfect agreement with previous works showing that the laser behaves as if the emitting dipoles were oriented along the crystallographic axes \cite{Schwartz2009}, although more elaborate models suggest more complex descriptions for the spectroscopy of Nd$^{3+}$ ions embedded in YAG matrix \cite{Dalgliesh1998, McKay2007}. 
This work opens interesting perspectives in several directions. First, it shows that a properly designed active medium with a careful control of the orientation of the emitting dipoles would permit to solve the same problem at other wavelengths, such as, e. g., at 1.5 $\mu$m where dual-frequency lasers can be used for LIDAR-RADAR applications \cite{Morvan2002} and for optical distribution of RF local oscillators through optical fibers \cite{Pillet2008}. Second, it opens new perspectives of applications of dual-frequency solid-state lasers, in domains in which the noises in the intensities in the two polarization modes play a central role, such as the probing of cesium clocks based on coherent population trapping phenomena \cite{Zanon2005}. Finally, this work shows that anti-phase relaxation oscillations can be an interesting probe to investigate the orientations of the emitting dipoles in solid-state laser media.

This research is partially funded by the Contrat de Projet Etat-R\'egion PONANT. The authors acknowledge Eric Collet and Lo\"\i c Toupet for the X-ray measurements and Cyril Hamel for the technical support.

\end{document}